\documentclass[aps,prl,twocolumn,showpacs,preprintnumbers,amsmath,amssymb]{revtex4}

\usepackage{color}
\usepackage{amsmath,amssymb}

\usepackage{color}
\usepackage[dvips]{graphicx}
\usepackage{subfigure}

\newcommand{\lsim}{\mathrel{\mathop{\kern 0pt \rlap
  {\raise.2ex\hbox{$<$}}}
  \lower.9ex\hbox{\kern-.190em $\sim$}}}
\newcommand{\gsim}{\mathrel{\mathop{\kern 0pt \rlap
  {\raise.2ex\hbox{$>$}}}
  \lower.9ex\hbox{\kern-.190em $\sim$}}}
\newcommand{\sigmav}{\langle \sigma_{\rm ann} v \rangle}

\newcommand{\pbar}{\bar{p}}

\newcommand{\beq}{\begin{equation}}
\newcommand{\eeq}{\end{equation}}
\newcommand{\bea}{\begin{eqnarray}}
\newcommand{\ena}{\end{eqnarray}}

\def\pbar{$\overline{p}$}

\def\apj{ApJ}%
\def\apjl{ApJ}%
\def\aap{A\&A}%
\def\plb{Phys.~Lett.~B}
\def\prd{Phys.~Rev.~D}%
\def\prl{Phys.~Rev.~Lett.}%
\def\ap{Astropart.~Phys.}%
\def\physrep{Phys.~Rep.}%
\begin{document}

\preprint{DFTT 28/2008, LAPTH-1273/08, IRFU-08-190}

\title{Constraints on WIMP Dark Matter from the High Energy PAMELA $\bar{p}/p$ data}

\author{F. Donato}
\affiliation{Dipartimento di Fisica Teorica, Universit\`a di Torino \\
Istituto Nazionale di Fisica Nucleare, via P. Giuria 1, I--10125 Torino, Italy}

\author{D. Maurin}
\affiliation{ Laboratoire de Physique Nucl\'eaire et Hautes Energies,
CNRS-IN2P3/Universit\'e Paris VII, 4 Place Jussieu, Tour 33, 75252 Paris Cedex
05, France}

\author{P. Brun}
\affiliation{CEA, Irfu, Service de Physique des Particules, Centre de Saclay, F-91191
Gif-sur-Yvette, France}

\author{T. Delahaye}
\affiliation{LAPTH, Universit\'e de Savoie, CNRS, B.P.110 74941 Annecy-le-Vieux, France}

\author{P. Salati}
\affiliation{LAPTH, Universit\'e de Savoie, CNRS, B.P.110 74941 Annecy-le-Vieux, France}

\date{\today}

\begin{abstract}
A new calculation of the $\bar{p}/p$ ratio in cosmic rays is compared to the
recent PAMELA data. The good match up to 100 GeV allows to
set constraints on exotic contributions from thermal WIMP dark matter candidates.
We derive stringent limits on possible enhancements of the WIMP \pbar\ flux:
a $m_{\rm WIMP}$=100 GeV  (1 TeV) signal cannot be increased by more than a
factor 6 (40) without overrunning PAMELA data. Annihilation through the $W^+W^-$
channel is also inspected and cross-checked with $e^+/(e^-+e^+)$ data. 
This scenario is strongly disfavored as it fails to simultaneously reproduce
positron and antiproton measurements.
\end{abstract}


\maketitle

The cosmic ray (CR) antiproton and positron fluxes
are considered as prime targets for indirect detection of galactic dark
matter (DM).
A deviation from the predicted astrophysical background has been searched
for mostly at low energy (e.g., \citep{1998PhRvD..58l3503B}).
However, in some scenario, heavy Weakly Interacting Massive Particle (WIMP)
candidates|either from annihilation
of the lightest supersymmetric species, or from the lightest
Kaluza-Klein particles in universal extra dimensions|should
be able to provide sizeable fluxes beyond a few GeV \citep{2007PhRvD..75h3006B}.
The PAMELA collaboration has recently published the cosmic ray
antiproton to proton ratio in the hitherto unexplored $\sim 1-100$
GeV energy range \cite{PAMELA_PRL}.
We find that the background flux, yielded by standard astrophysical processes,
can explain the data up to high energies. 
Adding a contribution from the annihilation of a generic WIMP dark matter halo, 
we derive stringent upper limits on possible boost factors of the exotic \pbar\ flux.
In addition, in some scenarios, the WIMPs mostly annihilate into $W^+W^-$ pairs,
hence giving rise to a copious amount of hard positrons. 
Recent calculations, in the light of the new PAMELA measurement of the
positronic fraction in CRs, showed that the secondary production alone
could marginally explain the data \cite{Delahaye:2008ua}. Should a heavy WIMP be
required to better match the positron flux, we also check the viability
of such models through the combined constraints from the $e^+$ and \pbar\
fluxes as also done in \cite{Cirelli:2008pk}.
\begin{figure*}[!t]
\includegraphics[width=\columnwidth]{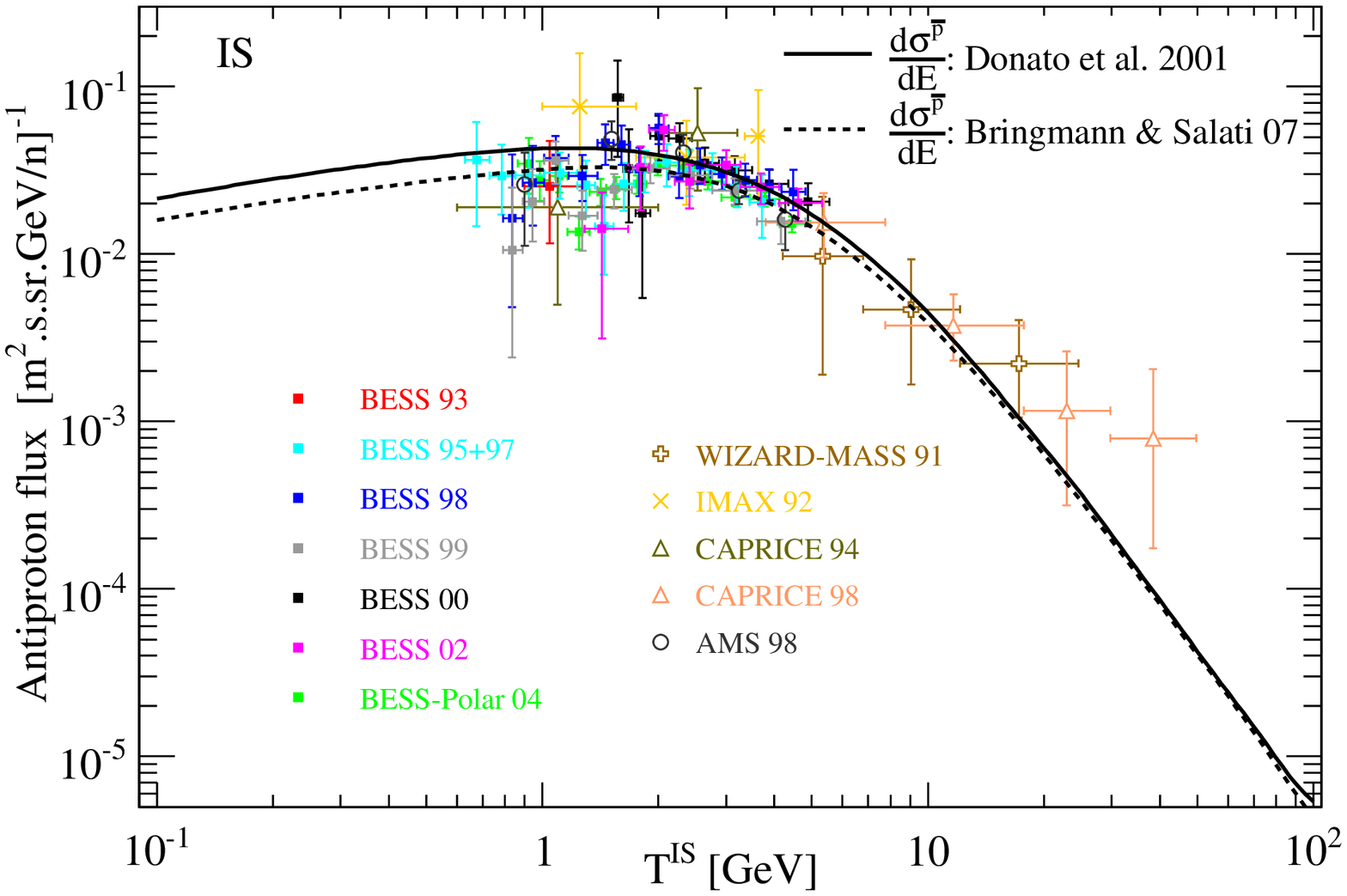}
\includegraphics[width=\columnwidth]{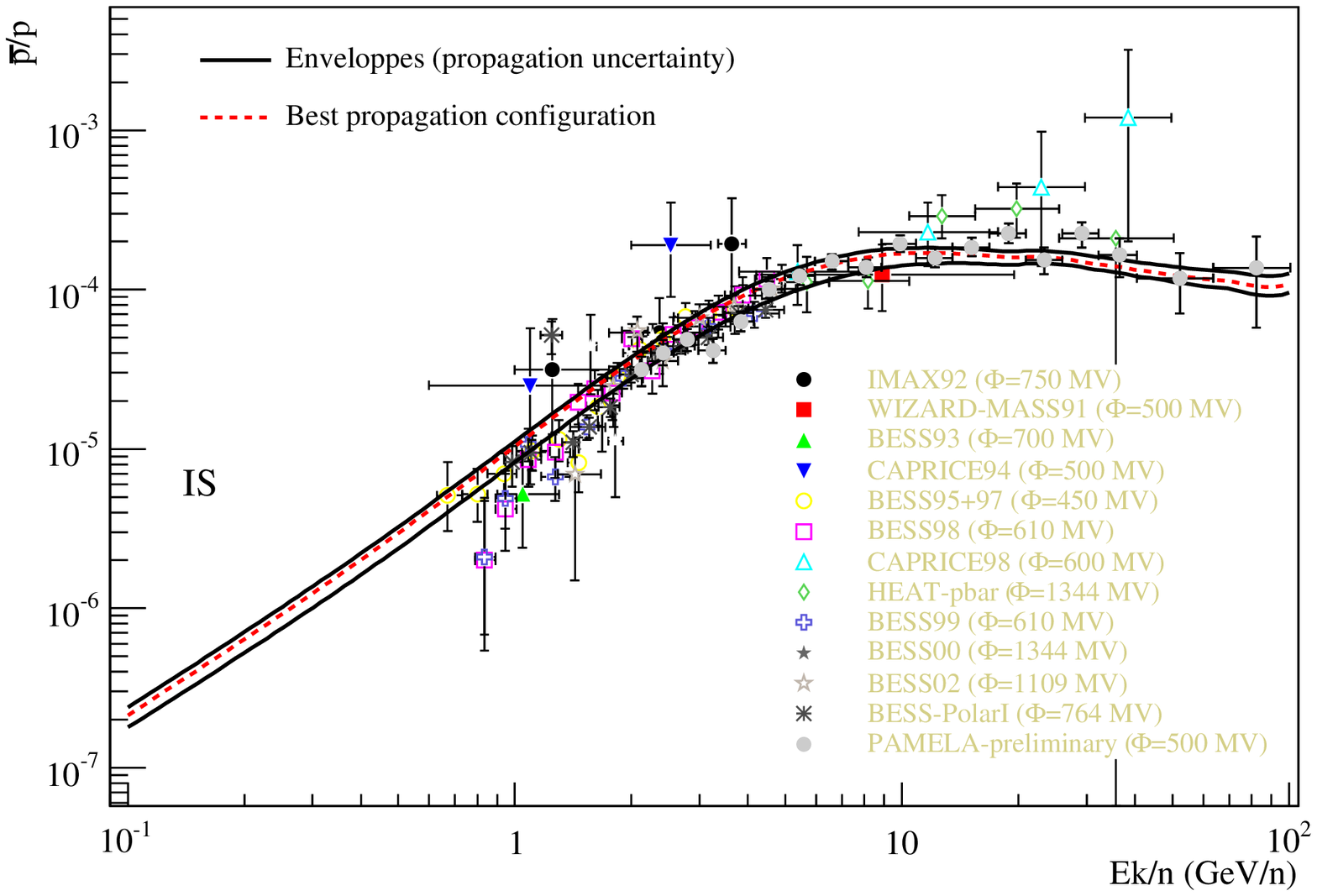}
\caption{Left panel: IS antiproton flux for the B/C best fit model and two
parameterizations of the production cross section.
Right panel: propagation uncertainty envelopes of the IS \pbar$/p$ ratio
for the same production cross sections as in the left panel.
All data are demodulated using the force-field approximation: AMS~98 \cite{2002PhR...366..331A},
IMAX~92 \cite{1996PhRvL..76.3057M}, CAPRICE~94 \cite{1997ApJ...487..415B},
WIZARD-MASS~91 \cite{1999ICRC....3...77B},  CAPRICE~98 \cite{2001ApJ...561..787B},
BESS93 \cite{1997ApJ...474..479M}, BESS~95+97 \cite{2000PhRvL..84.1078O},
BESS~98 \cite{2001APh....16..121M}, BESS~99 and 2000 \cite{2002PhRvL..88e1101A},
BESS~2002 \cite{2005ICRC....3...13H}, BESS Polar \cite{2008arXiv0805.1754A},
WIZARD-MASS~1~\cite{1996ApJ...467L..33H}, HEAT-$\bar{p}$ \cite{2001PhRvL..87A1101B},
and  PAMELA \cite{PAMELA_PRL}.}
\label{fig:pbar_sec}
\end{figure*}
The secondary \pbar\ flux provided in this paper is an improved
calculation of that presented in \cite{2001ApJ...563..172D},
to which we refer for a thorough
discussion of the ingredients and the technical details.
Antiprotons are yielded by the spallation of cosmic ray 
proton and helium nuclei over the interstellar medium, 
 the contribution of heavier nuclei being 
negligible \citep{2005PhRvD..71h3013D}.
Even if only p and He with kinetic energy larger than $6\,m_p$ can produce \pbar, 
a good description of the p and He interstellar (IS)
 fluxes is mandatory to correctly provide
the \pbar$/p$ ratio in the 0.1~GeV$-$100~GeV range.
Following \cite{2007APh....28..154S}, we model the proton and helium IS fluxes as 
\begin{equation}
  \Phi = A \beta^{P_1} R^{-P_2} \;\;{\rm m}^{-2}~{\rm s}^{-1}~{\rm sr}^{-1}~({\rm GeV/n})^{-1},
  \label{eq:flux_fit}
\end{equation}
where $R$ is the rigidity of the particle. 
The parameterization for the fluxes below 20 GeV/n are 
taken from the reanalysis of the 1997 to 2002 BESS data \cite{2007APh....28..154S}:
$\{A,\,P_1,\,P_2\}=\{19400,\,0.7,\,2.76\}$ for H and  $\{7100,\,0.5,\,2.78\}$ for He.
For the high energy range, the combined fit of AMS-01 \citep{2000PhLB..490...27A,2000PhLB..472..215A,2000PhLB..494..193A},
BESS98 \citep{2000ApJ...545.1135S} and BESS-TeV \citep{2004PhLB..594...35H} 
demodulated data respectively
give $\{A,\,P_1,\,P_2\}=\{24132,\,0.,\,2.84\}$ for H and  $\{8866,\,0.,\,2.85\}$ for He.
The two fits connect smoothly at 20~GeV/n. 
Compared to \citep{2001ApJ...563..172D}, we also
improve the calculation of the tertiary mechanism \citep{1999ApJ...526..215B}.
The Anderson prescription \citep{1967PhRvL..19..198A} is used, as described in
\citep{2005PhRvD..71h3013D,2008PhRvD..78d3506D}. As a result, the low
energy tail is more replenished, leading to a larger flux.
The framework used to calculate cosmic ray fluxes is the diffusion model with
convection and reacceleration. The transport parameters 
 are fixed from the boron-to-carbon (B/C) analysis \citep{2001ApJ...555..585M} and
correspond to i) the diffusion halo
of the Galaxy $L$; ii) the normalization of the diffusion coefficient $K_0$ and
its slope $\delta$ ($K(E)=K_0 \beta R^\delta$);
iii) the velocity of the constant wind directed perpendicular to the galactic disk
$\vec{V_c} = \pm V_c \vec{e_z}$; and iv) the reacceleration strength mediated via
the Alfv\'enic speed $V_a$. Strong degeneracies are observed among the allowed parameter
sets \citep{2001ApJ...555..585M}, but it has a limited impact on the secondary \pbar\ flux
\citep{2001ApJ...563..172D}. At variance, the corresponding DM-induced \pbar\ flux suffers
large propagation uncertainties \citep{2004PhRvD..69f3501D}; this is also the case 
for the secondary and primary positron fluxes \citep{2008PhRvD..77f3527D,Delahaye:2008ua}.
Throughout the paper, the fluxes will be shown for the B/C  best fit propagation parameters,
i.e. $L = 4.$~kpc, $K_0 = 0.0112$~kpc$^2$Myr$^{-1}$, $\delta=0.7$,
$V_c = 12.$~km~s$^{-1}$ and $V_a = 52.9$~km~s$^{-1}$ \cite{2001ApJ...555..585M}.

 The secondary IS \pbar\ flux is displayed in the left panel of Fig.~\ref{fig:pbar_sec}
along with the data demodulated according to the force-field prescription.
We either use the DTUNUC \cite{2001ApJ...563..172D} \pbar\ production cross sections
(solid line) or those discussed in \cite{2005PhRvD..71h3013D,2007PhRvD..75h3006B} (dashed line).
The differences between the two curves illustrate the uncertainty related to the production
cross sections, as emphasized in \cite{2001ApJ...563..172D}, where a careful
and conservative analysis within the DTUNUC simulation settled a nuclear 
uncertainty of $\sim 25\%$ over the energy range $0.1-100$~GeV.
The conclusion is similar here, although the two sets of cross sections differ
mostly at low energy.
In the right panel,  along with the demodulated \pbar$/p$ data, we show the curves
bounding the propagation uncertainty on the \pbar\ calculation based either on the
DTUNUC \cite{2001ApJ...563..172D} \pbar\ production cross sections (solid lines)
or those borrowed from \cite{2007PhRvD..75h3006B} (dashed lines). The uncertainty
arising from propagation is comparable to the nuclear one \cite{2001ApJ...563..172D}.
 From a bare eye inspection, it is evident that the secondary contribution alone explains
PAMELA data on the whole energetic range. It is not necessary to invoke an additional component
to the standard astrophysical one.

Motivated by the accuracy of our predictions and their 
 well understood theoretical uncertainties, as well as by the good statistical 
significance of PAMELA data, we derive limits on a possible exotic component.
We focus on the high energy part of the $\bar{p}/p$ ratio, where solar modulation does
not play any role \cite{1999PhRvL..83..674B}.
We assume an additional component of 
antiprotons produced by annihilation of WIMPs filling the dark
halo of the Milky Way. Their distribution is taken as a cored-isothermal sphere
with local density  $\rho_{\odot} = 0.3$ GeV cm$^{-3}$. The velocity-averaged
annihilation cross section is taken as $\sigmav = 3 \times 10^{-26}$ cm$^{3}$s$^{-1}$,
with an annihilation channel into $b$-$\bar{b}$. According to \cite{2004PhRvD..69f3501D},
the propagated primary \pbar\ flux is only very mildly dependent on the annihilation
channel and the DM distribution function. Therefore, our assumptions can be considered
valid for a generic  WIMP dark matter candidate except for a rough rescaling factor. 
Propagation is treated in the same way as for the secondary component
\cite{2004PhRvD..69f3501D,2008PhRvD..78d3506D}. 
As a reference case, we employ the best fit transport parameters listed above and recall
that the uncertainty on the primary \pbar\ flux due to propagation spans roughly one order
of magnitude above and one below the best fit scenario. 
\begin{figure}[t]
\includegraphics[width=1.\columnwidth]{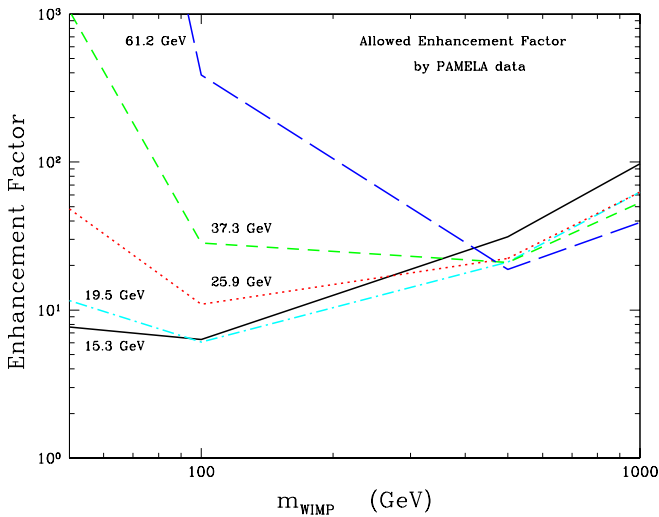}
\caption{Upper limits on the enhancement factor to the primary \pbar\ flux as a function
of the WIMP mass, derived from a comparison with PAMELA high energy data. Each curve
is labelled according to the corresponding PAMELA energy bin.}
\label{fig:boost}
\end{figure}
We add the calculated primary \pbar\ flux for different WIMP
masses to the secondary component and compare the total flux to PAMELA 
high energy data, namely $T_{\bar{p}}> $ 10 GeV. To be conservative,
the background calculated from Bringmann \& Salati's \pbar\ production
cross sections is considered (dashed curves in Fig.~\ref{fig:pbar_sec}).
We derive the factor by which the DM flux could be enhanced without exceeding
experimental data (2$\sigma$ error bars) in any energy bin.
The maximum allowed enhancement factor is 
plotted in Fig.~\ref{fig:boost} as a function of the WIMP mass: 
it cannot exceed 6--20--40 for  $m_{\rm WIMP}$=100--500--1000 GeV, respectively. 
These limits can be reinforced as well as relaxed by quite simple 
modifications of the key ingredients in the flux calculation,
just as described above. 
The boost factor may be ascribed, in principle, to clumpiness in the 
DM distribution \citep{2008A&A...479..427L}|this contribution being
energy-dependent|as well as to an increase 
of the annihilation cross section as proposed by
\cite{1989PhLB..225..372B} and more recently by
\cite{Hisano:2003ec} using the Sommerfeld effect.
\begin{figure*}[t!]
\includegraphics[width=\columnwidth]{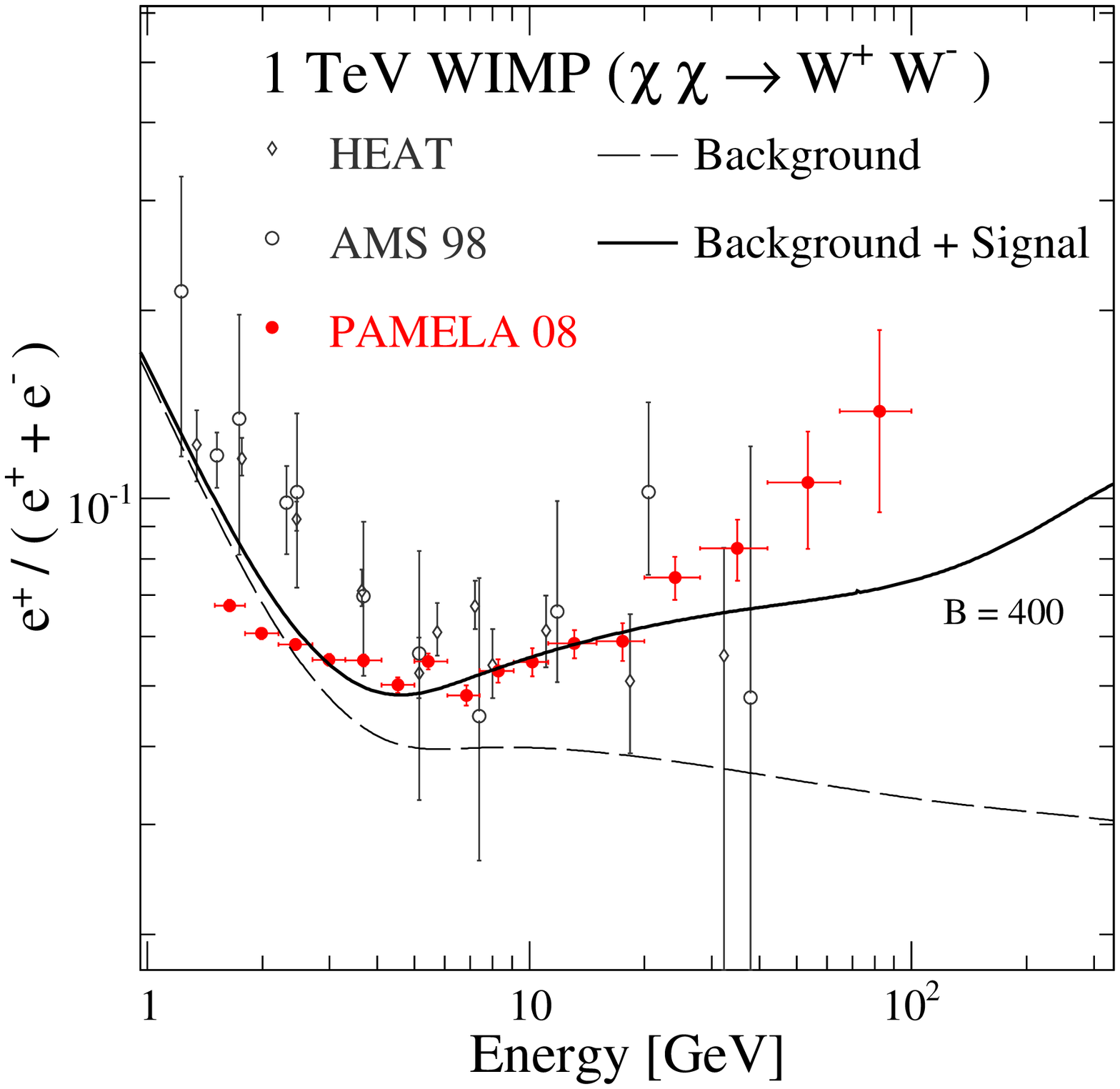}
\includegraphics[width=0.96\columnwidth]{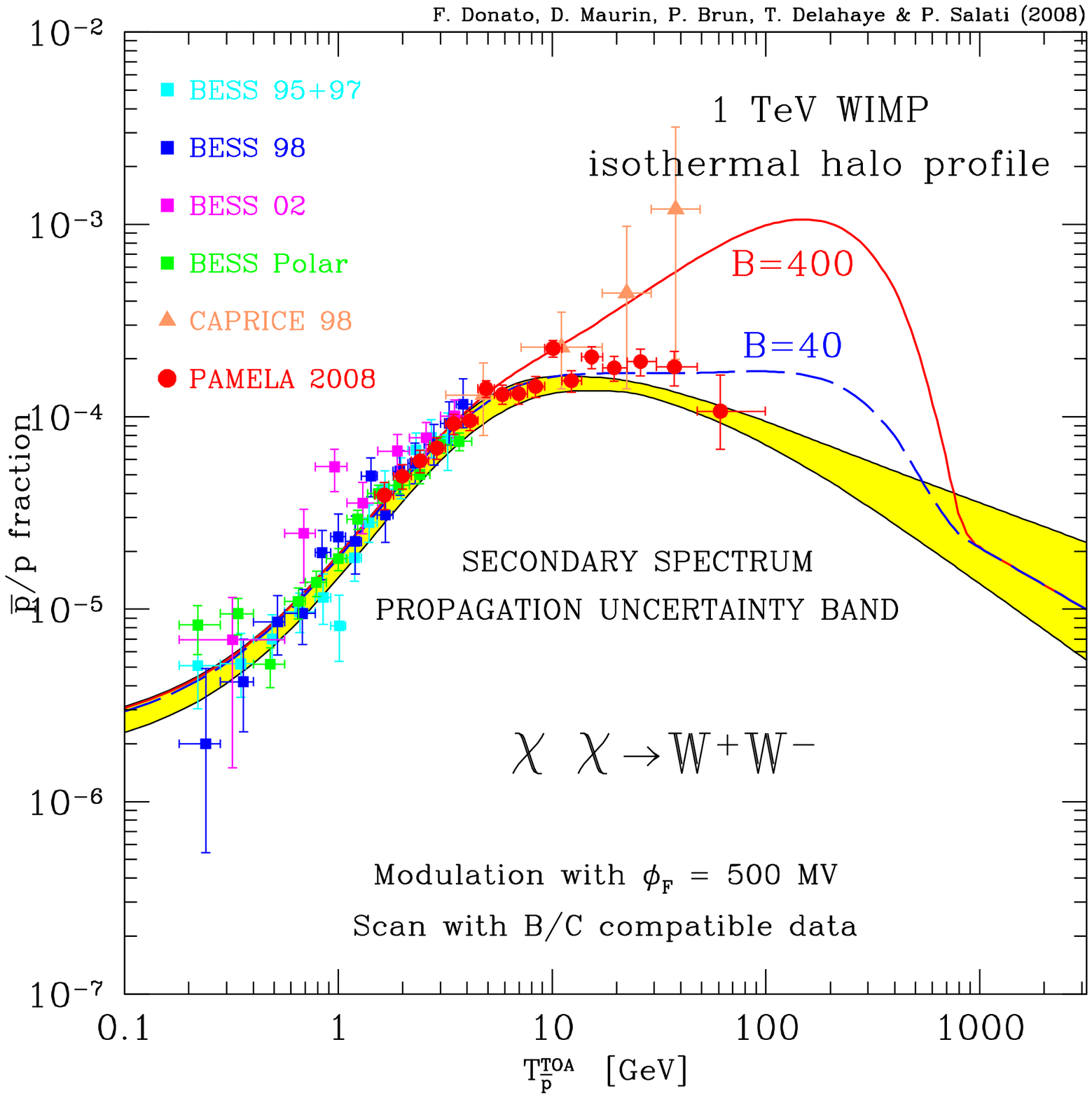}
\caption{
The fiducial case of a 1 TeV LSP annihilating into a $W^{+} W^{-}$ pair
is featured. In the left panel, the positron signal which this DM species
yields has been increased by a factor of 400, hence the solid curve and
a marginal agreement with the PAMELA data. 
Positron fraction data are from HEAT \cite{1997ApJ...482L.191B}, 
AMS-01 \cite{Aguilar:2007yf,Alcaraz:2000bf} and PAMELA \cite{PAMELA_frac}.
If the so--called Sommerfeld effect \cite{Hisano:2003ec} is invoked to
explain such a large enhancement of the annihilation cross section, the
same boost applies to antiprotons and leads to an unacceptable distortion
of their spectrum as indicated by the red solid line of the right panel.
}
\label{fig:1_TEV_WIMP}
\end{figure*}

Our conclusions have important consequences on the explanations of the
positron data based on the annihilation of DM species within
the Milky Way halo.
The positron fraction suffers from large uncertainties related for
instance to the poorly determined electron spectral index above
10~GeV~\cite{Delahaye:2008ua}. Although soft electrons are associated
to large values of the positron fraction and to a marginal agreement of the
pure secondary positron flux with the measurements, we cannot dismiss
the possibility of a hard cosmic ray electron distribution. A spectral
index of 3.44 \cite{2004ApJ...612..262C} leads actually in the left panel of Fig.~\ref{fig:1_TEV_WIMP}
to the long-dashed curve featuring a low background case.
With a typical annihilation cross section $\sigmav$ of
$3 \times 10^{-26}$ cm$^{3}$s$^{-1}$, WIMPs do not produce enough positrons to
reproduce the increasing trend observed in 
$e^+/(e^+ + e^-)$ data \cite{PAMELA_frac},so that a significant
enhancement of the annihilation rate is necessary as shown
in~\cite{Bergstrom:2008gr}. However the boost factor associated
to DM clumps cannot exceed at most a factor of $\sim$ 10 in the standard
$\Lambda$-CDM scenario of structure formation~\cite{2008A&A...479..427L}.
Astrophysics does not provide then a natural explanation for the large
boost factors required to fit the positron excess. That is why the Sommerfeld
effect~\cite{Hisano:2003ec} has been advocated as a plausible mechanism
to significantly increase the WIMP annihilation cross section in the
non-relativistic regime prevailing today in galactic haloes. Heavy
DM species is a prerequisite. We then consider a generic 1 TeV
particle annihilating into $W^{+} W^{-}$ pairs and 
boost $\sigmav$ by a factor of 400 in order to get the solid line in
the left panel of Fig.~\ref{fig:1_TEV_WIMP}. Although an annihilation
cross section of $1.2 \times 10^{-23}$ cm$^{3}$s$^{-1}$ is possible should
non-perturbative effects be involved, the consequences on antiprotons
are drastic. The red solid curve in the right panel of
Fig.~\ref{fig:1_TEV_WIMP} features an unacceptable distortion of the
\pbar\ spectrum. The DM positron signal cannot be enhanced 
without playing havoc with the \pbar\ measurements. 

Nonetheless, notice that ways out are possible whose careful investigation is
beyond the scope of this Letter.
The value of 400 assumed for the positron signal of
Fig.~\ref{fig:1_TEV_WIMP} could arise from the combined effects of DM
clumpiness and $\sigmav$ enhancement. If a generous factor of 10
is assumed for the former|a marginally acceptable
value~\cite{2008A&A...479..427L}|the latter does not exceed 40.
Unlike positrons which are produced locally, the antiprotons detected at
the Earth originate from a large region of the Milky Way halo over which
substructures may not be as important as in our vicinity. The \pbar\ flux
may not be much enhanced by the presence of DM clumps so that a value of 40
would apply in that case to the antiproton boost. The corresponding blue
long-dashed line in the right panel of Fig.~\ref{fig:1_TEV_WIMP}
features a fairly acceptable \pbar\ spectrum.

Viable scenarios such as a large black hole population pervading the Galaxy
\cite{Bertone:2005xz,Brun:2007tn} also lead to large boost factors although
it seems difficult a priori to prevent the antiproton flux from being enhanced
too much.
Notice finally that the cosmic ray propagation model could be different from the
one selected here. Once again, positron and antiproton fluxes have
different behaviors toward a change in the propagation parameters.
For example, the primary \pbar\ flux could be easily decreased by
an order of magnitude without violating B/C data, allowing a Sommerfeld
boost of the cross section of 400.

A new calculation for the secondary cosmic antiproton flux
 and the relevant uncertainties have been presented.
The ratio $\bar{p}/p$ has been derived after fitting recent proton data. 
Our predictions can explain the experimental data, and 
in particular the recent PAMELA data, which span more than two
decades in energy. No exotic contribution|as from annihilating 
dark matter in the galactic halo|has to be invoked to reproduce 
experimental results. 
Analyzing the high energy part of the PAMELA $\bar{p}/p$ we derive 
strong upper limits on possible enhancements of the exotic 
\pbar\ flux as a function of the WIMP mass. 
Relying on standard assumptions the exotic antiproton flux induced by a $m_{\rm WIMP}=100$ GeV  
(1 TeV) DM halo cannot be 
increased by more than a factor 6 (40) without overrunning PAMELA data.
Would the Sommerfeld effect  ($W^+W^-$ channel) be invoked to
explain PAMELA leptonic data, the corresponding enhancement of the \pbar\ production
would lead to an unacceptable distortion of the $\bar{p}/p$ spectrum.
\\
{\it Acknowledgements.} F.D. thanks N. Fornengo and M. Boezio for 
useful discussions. We acknowledge the support from the French Programme
National de Cosmologie and the Explora'doc PhD student program.


\begin{thebibliography}{44}
\expandafter\ifx\csname natexlab\endcsname\relax\def\natexlab#1{#1}\fi
\expandafter\ifx\csname bibnamefont\endcsname\relax
  \def\bibnamefont#1{#1}\fi
\expandafter\ifx\csname bibfnamefont\endcsname\relax
  \def\bibfnamefont#1{#1}\fi
\expandafter\ifx\csname citenamefont\endcsname\relax
  \def\citenamefont#1{#1}\fi
\expandafter\ifx\csname url\endcsname\relax
  \def\url#1{\texttt{#1}}\fi
\expandafter\ifx\csname urlprefix\endcsname\relax\def\urlprefix{URL }\fi
\providecommand{\bibinfo}[2]{#2}
\providecommand{\eprint}[2][]{\url{#2}}

\bibitem[{\citenamefont{{Bottino} et~al.}(1998)\citenamefont{{Bottino},
  {Donato}, {Fornengo}, and {Salati}}}]{1998PhRvD..58l3503B}
\bibinfo{author}{\bibfnamefont{A.}~\bibnamefont{{Bottino}}},
  \bibinfo{author}{\bibfnamefont{F.}~\bibnamefont{{Donato}}},
  \bibinfo{author}{\bibfnamefont{N.}~\bibnamefont{{Fornengo}}},
  \bibnamefont{and} \bibinfo{author}{\bibfnamefont{P.}~\bibnamefont{{Salati}}},
  \bibinfo{journal}{\prd} \textbf{\bibinfo{volume}{58}},
  \bibinfo{pages}{123503} (\bibinfo{year}{1998}).

\bibitem[{\citenamefont{{Bringmann} and {Salati}}(2007)}]{2007PhRvD..75h3006B}
\bibinfo{author}{\bibfnamefont{T.}~\bibnamefont{{Bringmann}}} \bibnamefont{and}
  \bibinfo{author}{\bibfnamefont{P.}~\bibnamefont{{Salati}}},
  \bibinfo{journal}{\prd} \textbf{\bibinfo{volume}{75}},
  \bibinfo{pages}{083006} (\bibinfo{year}{2007}).

\bibitem[{\citenamefont{{Adriani} et~al.}(2009)}]{PAMELA_PRL}
\bibinfo{author}{\bibfnamefont{O.}~\bibnamefont{{Adriani}}}
  \bibnamefont{et~al.} (\bibinfo{collaboration}{PAMELA}),
  \bibinfo{journal}{\prl} \textbf{\bibinfo{volume}{102}},
  \bibinfo{pages}{051101} (\bibinfo{year}{2009}).

\bibitem[{\citenamefont{{Delahaye}
  et~al.}(2008{\natexlab{a}})}]{Delahaye:2008ua}
\bibinfo{author}{\bibfnamefont{T.}~\bibnamefont{{Delahaye}}}
  \bibnamefont{et~al.}, \bibinfo{journal}{\aap\ submitted}
  (\bibinfo{year}{2008}{\natexlab{a}}), \eprint{0809.5268}.

\bibitem[{\citenamefont{{Cirelli} et~al.}(2008)\citenamefont{{Cirelli},
  {Kadastik}, {Raidal}, and {Strumia}}}]{Cirelli:2008pk}
\bibinfo{author}{\bibfnamefont{M.}~\bibnamefont{{Cirelli}}},
  \bibinfo{author}{\bibfnamefont{M.}~\bibnamefont{{Kadastik}}},
  \bibinfo{author}{\bibfnamefont{M.}~\bibnamefont{{Raidal}}}, \bibnamefont{and}
  \bibinfo{author}{\bibfnamefont{A.}~\bibnamefont{{Strumia}}}
  (\bibinfo{year}{2008}), \eprint{0809.2409}.

\bibitem[{\citenamefont{{Aguilar} et~al.}(2002)}]{2002PhR...366..331A}
\bibinfo{author}{\bibfnamefont{M.}~\bibnamefont{{Aguilar}}}
  \bibnamefont{et~al.} (\bibinfo{collaboration}{AMS}),
  \bibinfo{journal}{\physrep} \textbf{\bibinfo{volume}{366}},
  \bibinfo{pages}{331} (\bibinfo{year}{2002}).

\bibitem[{\citenamefont{{Mitchell} et~al.}(1996)}]{1996PhRvL..76.3057M}
\bibinfo{author}{\bibfnamefont{J.~W.} \bibnamefont{{Mitchell}}}
  \bibnamefont{et~al.} (\bibinfo{collaboration}{IMAX}), \bibinfo{journal}{\prl}
  \textbf{\bibinfo{volume}{76}}, \bibinfo{pages}{3057} (\bibinfo{year}{1996}).

\bibitem[{\citenamefont{{Boezio} et~al.}(1997)}]{1997ApJ...487..415B}
\bibinfo{author}{\bibfnamefont{M.}~\bibnamefont{{Boezio}}} \bibnamefont{et~al.}
  (\bibinfo{collaboration}{CAPRICE}), \bibinfo{journal}{\apj}
  \textbf{\bibinfo{volume}{487}}, \bibinfo{pages}{415} (\bibinfo{year}{1997}).

\bibitem[{\citenamefont{{Basini} et~al.}(1999)}]{1999ICRC....3...77B}
\bibinfo{author}{\bibfnamefont{G.}~\bibnamefont{{Basini}}} \bibnamefont{et~al.}
  (\bibinfo{collaboration}{WIZARD-MASS}), \bibinfo{journal}{ICRC}
  \textbf{\bibinfo{volume}{3}}, \bibinfo{pages}{77} (\bibinfo{year}{1999}).

\bibitem[{\citenamefont{{Boezio} et~al.}(2001)}]{2001ApJ...561..787B}
\bibinfo{author}{\bibfnamefont{M.}~\bibnamefont{{Boezio}}} \bibnamefont{et~al.}
  (\bibinfo{collaboration}{CAPRICE}), \bibinfo{journal}{\apj}
  \textbf{\bibinfo{volume}{561}}, \bibinfo{pages}{787} (\bibinfo{year}{2001}).

\bibitem[{\citenamefont{{Moiseev} et~al.}(1997)}]{1997ApJ...474..479M}
\bibinfo{author}{\bibfnamefont{A.}~\bibnamefont{{Moiseev}}}
  \bibnamefont{et~al.} (\bibinfo{collaboration}{BESS}), \bibinfo{journal}{\apj}
  \textbf{\bibinfo{volume}{474}}, \bibinfo{pages}{479} (\bibinfo{year}{1997}).

\bibitem[{\citenamefont{{Orito} et~al.}(2000)}]{2000PhRvL..84.1078O}
\bibinfo{author}{\bibfnamefont{S.}~\bibnamefont{{Orito}}} \bibnamefont{et~al.}
  (\bibinfo{collaboration}{BESS}), \bibinfo{journal}{\prl}
  \textbf{\bibinfo{volume}{84}}, \bibinfo{pages}{1078} (\bibinfo{year}{2000}).

\bibitem[{\citenamefont{{Maeno} et~al.}(2001)}]{2001APh....16..121M}
\bibinfo{author}{\bibfnamefont{T.}~\bibnamefont{{Maeno}}} \bibnamefont{et~al.}
  (\bibinfo{collaboration}{BESS}), \bibinfo{journal}{\ap}
  \textbf{\bibinfo{volume}{16}}, \bibinfo{pages}{121} (\bibinfo{year}{2001}).

\bibitem[{\citenamefont{{Asaoka} et~al.}(2002)}]{2002PhRvL..88e1101A}
\bibinfo{author}{\bibfnamefont{Y.}~\bibnamefont{{Asaoka}}} \bibnamefont{et~al.}
  (\bibinfo{collaboration}{BESS}), \bibinfo{journal}{\prl}
  \textbf{\bibinfo{volume}{88}}, \bibinfo{pages}{051101}
  (\bibinfo{year}{2002}).

\bibitem[{\citenamefont{{Haino} et~al.}(2005)}]{2005ICRC....3...13H}
\bibinfo{author}{\bibfnamefont{S.}~\bibnamefont{{Haino}}} \bibnamefont{et~al.}
  (\bibinfo{collaboration}{BESS}), \bibinfo{journal}{ICRC}
  \textbf{\bibinfo{volume}{3}}, \bibinfo{pages}{13} (\bibinfo{year}{2005}).

\bibitem[{\citenamefont{{Abe} et~al.}(2008)}]{2008arXiv0805.1754A}
\bibinfo{author}{\bibfnamefont{K.}~\bibnamefont{{Abe}}} \bibnamefont{et~al.}
  (\bibinfo{collaboration}{BESS}), \bibinfo{journal}{ArXiv e-prints}
  (\bibinfo{year}{2008}), \eprint{0805.1754}.

\bibitem[{\citenamefont{{Hof} et~al.}(1996)}]{1996ApJ...467L..33H}
\bibinfo{author}{\bibfnamefont{M.}~\bibnamefont{{Hof}}} \bibnamefont{et~al.}
  (\bibinfo{collaboration}{WIZARD-MASS 1}), \bibinfo{journal}{\apjl}
  \textbf{\bibinfo{volume}{467}}, \bibinfo{pages}{L33} (\bibinfo{year}{1996}).

\bibitem[{\citenamefont{{Beach} et~al.}(2001)}]{2001PhRvL..87A1101B}
\bibinfo{author}{\bibfnamefont{A.~S.} \bibnamefont{{Beach}}}
  \bibnamefont{et~al.} (\bibinfo{collaboration}{HEAT}), \bibinfo{journal}{\prl}
  \textbf{\bibinfo{volume}{87}}, \bibinfo{pages}{271101}
  (\bibinfo{year}{2001}).

\bibitem[{\citenamefont{{Donato} et~al.}(2001)\citenamefont{{Donato}, {Maurin},
  {Salati}, {Barrau}, {Boudoul}, and {Taillet}}}]{2001ApJ...563..172D}
\bibinfo{author}{\bibfnamefont{F.}~\bibnamefont{{Donato}}},
  \bibinfo{author}{\bibfnamefont{D.}~\bibnamefont{{Maurin}}},
  \bibinfo{author}{\bibfnamefont{P.}~\bibnamefont{{Salati}}},
  \bibinfo{author}{\bibfnamefont{A.}~\bibnamefont{{Barrau}}},
  \bibinfo{author}{\bibfnamefont{G.}~\bibnamefont{{Boudoul}}},
  \bibnamefont{and}
  \bibinfo{author}{\bibfnamefont{R.}~\bibnamefont{{Taillet}}},
  \bibinfo{journal}{\apj} \textbf{\bibinfo{volume}{563}}, \bibinfo{pages}{172}
  (\bibinfo{year}{2001}).

\bibitem[{\citenamefont{{Duperray} et~al.}(2005)}]{2005PhRvD..71h3013D}
\bibinfo{author}{\bibfnamefont{R.}~\bibnamefont{{Duperray}}}
  \bibnamefont{et~al.}, \bibinfo{journal}{\prd} \textbf{\bibinfo{volume}{71}},
  \bibinfo{pages}{083013} (\bibinfo{year}{2005}).

\bibitem[{\citenamefont{{Shikaze} et~al.}(2007)}]{2007APh....28..154S}
\bibinfo{author}{\bibfnamefont{Y.}~\bibnamefont{{Shikaze}}}
  \bibnamefont{et~al.} (\bibinfo{collaboration}{BESS}), \bibinfo{journal}{\ap}
  \textbf{\bibinfo{volume}{28}}, \bibinfo{pages}{154} (\bibinfo{year}{2007}).

\bibitem[{\citenamefont{{Alcaraz}
  et~al.}(2000{\natexlab{a}})}]{2000PhLB..490...27A}
\bibinfo{author}{\bibfnamefont{J.}~\bibnamefont{{Alcaraz}}}
  \bibnamefont{et~al.} (\bibinfo{collaboration}{AMS}), \bibinfo{journal}{\plb}
  \textbf{\bibinfo{volume}{490}}, \bibinfo{pages}{27}
  (\bibinfo{year}{2000}{\natexlab{a}}).

\bibitem[{\citenamefont{{Alcaraz}
  et~al.}(2000{\natexlab{b}})}]{2000PhLB..472..215A}
\bibinfo{author}{\bibfnamefont{J.}~\bibnamefont{{Alcaraz}}}
  \bibnamefont{et~al.} (\bibinfo{collaboration}{AMS}), \bibinfo{journal}{\plb}
  \textbf{\bibinfo{volume}{472}}, \bibinfo{pages}{215}
  (\bibinfo{year}{2000}{\natexlab{b}}).

\bibitem[{\citenamefont{{Alcaraz}
  et~al.}(2000{\natexlab{c}})}]{2000PhLB..494..193A}
\bibinfo{author}{\bibfnamefont{J.}~\bibnamefont{{Alcaraz}}}
  \bibnamefont{et~al.} (\bibinfo{collaboration}{AMS}), \bibinfo{journal}{\plb}
  \textbf{\bibinfo{volume}{494}}, \bibinfo{pages}{193}
  (\bibinfo{year}{2000}{\natexlab{c}}).

\bibitem[{\citenamefont{{Sanuki} et~al.}(2000)}]{2000ApJ...545.1135S}
\bibinfo{author}{\bibfnamefont{T.}~\bibnamefont{{Sanuki}}} \bibnamefont{et~al.}
  (\bibinfo{collaboration}{BESS}), \bibinfo{journal}{\apj}
  \textbf{\bibinfo{volume}{545}}, \bibinfo{pages}{1135} (\bibinfo{year}{2000}).

\bibitem[{\citenamefont{{Haino} et~al.}(2004)}]{2004PhLB..594...35H}
\bibinfo{author}{\bibfnamefont{S.}~\bibnamefont{{Haino}}} \bibnamefont{et~al.}
  (\bibinfo{collaboration}{BESS}), \bibinfo{journal}{\plb}
  \textbf{\bibinfo{volume}{594}}, \bibinfo{pages}{35} (\bibinfo{year}{2004}).

\bibitem[{\citenamefont{{Bergstr{\"o}m}
  et~al.}(1999)\citenamefont{{Bergstr{\"o}m}, {Edsj{\"o}}, and
  {Ullio}}}]{1999ApJ...526..215B}
\bibinfo{author}{\bibfnamefont{L.}~\bibnamefont{{Bergstr{\"o}m}}},
  \bibinfo{author}{\bibfnamefont{J.}~\bibnamefont{{Edsj{\"o}}}},
  \bibnamefont{and} \bibinfo{author}{\bibfnamefont{P.}~\bibnamefont{{Ullio}}},
  \bibinfo{journal}{\apj} \textbf{\bibinfo{volume}{526}}, \bibinfo{pages}{215}
  (\bibinfo{year}{1999}).

\bibitem[{\citenamefont{{Anderson} et~al.}(1967)}]{1967PhRvL..19..198A}
\bibinfo{author}{\bibfnamefont{E.~W.} \bibnamefont{{Anderson}}}
  \bibnamefont{et~al.}, \bibinfo{journal}{\prl} \textbf{\bibinfo{volume}{19}},
  \bibinfo{pages}{198} (\bibinfo{year}{1967}).

\bibitem[{\citenamefont{{Donato} et~al.}(2008)\citenamefont{{Donato},
  {Fornengo}, and {Maurin}}}]{2008PhRvD..78d3506D}
\bibinfo{author}{\bibfnamefont{F.}~\bibnamefont{{Donato}}},
  \bibinfo{author}{\bibfnamefont{N.}~\bibnamefont{{Fornengo}}},
  \bibnamefont{and} \bibinfo{author}{\bibfnamefont{D.}~\bibnamefont{{Maurin}}},
  \bibinfo{journal}{\prd} \textbf{\bibinfo{volume}{78}},
  \bibinfo{pages}{043506} (\bibinfo{year}{2008}).

\bibitem[{\citenamefont{{Maurin} et~al.}(2001)\citenamefont{{Maurin}, {Donato},
  {Taillet}, and {Salati}}}]{2001ApJ...555..585M}
\bibinfo{author}{\bibfnamefont{D.}~\bibnamefont{{Maurin}}},
  \bibinfo{author}{\bibfnamefont{F.}~\bibnamefont{{Donato}}},
  \bibinfo{author}{\bibfnamefont{R.}~\bibnamefont{{Taillet}}},
  \bibnamefont{and} \bibinfo{author}{\bibfnamefont{P.}~\bibnamefont{{Salati}}},
  \bibinfo{journal}{\apj} \textbf{\bibinfo{volume}{555}}, \bibinfo{pages}{585}
  (\bibinfo{year}{2001}).

\bibitem[{\citenamefont{{Donato} et~al.}(2004)\citenamefont{{Donato},
  {Fornengo}, {Maurin}, {Salati}, and {Taillet}}}]{2004PhRvD..69f3501D}
\bibinfo{author}{\bibfnamefont{F.}~\bibnamefont{{Donato}}},
  \bibinfo{author}{\bibfnamefont{N.}~\bibnamefont{{Fornengo}}},
  \bibinfo{author}{\bibfnamefont{D.}~\bibnamefont{{Maurin}}},
  \bibinfo{author}{\bibfnamefont{P.}~\bibnamefont{{Salati}}}, \bibnamefont{and}
  \bibinfo{author}{\bibfnamefont{R.}~\bibnamefont{{Taillet}}},
  \bibinfo{journal}{\prd} \textbf{\bibinfo{volume}{69}},
  \bibinfo{pages}{063501} (\bibinfo{year}{2004}).

\bibitem[{\citenamefont{{Delahaye}
  et~al.}(2008{\natexlab{b}})\citenamefont{{Delahaye}, {Lineros}, {Donato},
  {Fornengo}, and {Salati}}}]{2008PhRvD..77f3527D}
\bibinfo{author}{\bibfnamefont{T.}~\bibnamefont{{Delahaye}}},
  \bibinfo{author}{\bibfnamefont{R.}~\bibnamefont{{Lineros}}},
  \bibinfo{author}{\bibfnamefont{F.}~\bibnamefont{{Donato}}},
  \bibinfo{author}{\bibfnamefont{N.}~\bibnamefont{{Fornengo}}},
  \bibnamefont{and} \bibinfo{author}{\bibfnamefont{P.}~\bibnamefont{{Salati}}},
  \bibinfo{journal}{\prd} \textbf{\bibinfo{volume}{77}},
  \bibinfo{pages}{063527} (\bibinfo{year}{2008}{\natexlab{b}}),
  \eprint{0712.2312}.

\bibitem[{\citenamefont{{Bieber} et~al.}(1999)}]{1999PhRvL..83..674B}
\bibinfo{author}{\bibfnamefont{J.~W.} \bibnamefont{{Bieber}}}
  \bibnamefont{et~al.}, \bibinfo{journal}{\prl} \textbf{\bibinfo{volume}{83}},
  \bibinfo{pages}{674} (\bibinfo{year}{1999}).

\bibitem[{\citenamefont{{Lavalle} et~al.}(2008)\citenamefont{{Lavalle}, {Yuan},
  {Maurin}, and {Bi}}}]{2008A&A...479..427L}
\bibinfo{author}{\bibfnamefont{J.}~\bibnamefont{{Lavalle}}},
  \bibinfo{author}{\bibfnamefont{Q.}~\bibnamefont{{Yuan}}},
  \bibinfo{author}{\bibfnamefont{D.}~\bibnamefont{{Maurin}}}, \bibnamefont{and}
  \bibinfo{author}{\bibfnamefont{X.-J.} \bibnamefont{{Bi}}},
  \bibinfo{journal}{\aap} \textbf{\bibinfo{volume}{479}}, \bibinfo{pages}{427}
  (\bibinfo{year}{2008}), \eprint{0709.3634}.

\bibitem[{\citenamefont{{Bergstr{\"o}m}}(1989)}]{1989PhLB..225..372B}
\bibinfo{author}{\bibfnamefont{L.}~\bibnamefont{{Bergstr{\"o}m}}},
  \bibinfo{journal}{Physics Letters B} \textbf{\bibinfo{volume}{225}},
  \bibinfo{pages}{372} (\bibinfo{year}{1989}).

\bibitem[{\citenamefont{{Hisano} et~al.}(2004)\citenamefont{{Hisano},
  {Matsumoto}, and {Nojiri}}}]{Hisano:2003ec}
\bibinfo{author}{\bibfnamefont{J.}~\bibnamefont{{Hisano}}},
  \bibinfo{author}{\bibfnamefont{S.}~\bibnamefont{{Matsumoto}}},
  \bibnamefont{and} \bibinfo{author}{\bibfnamefont{M.~M.}
  \bibnamefont{{Nojiri}}}, \bibinfo{journal}{Phys. Rev. Lett.}
  \textbf{\bibinfo{volume}{92}}, \bibinfo{pages}{031303}
  (\bibinfo{year}{2004}).

\bibitem[{\citenamefont{{Barwick} et~al.}(1997)}]{1997ApJ...482L.191B}
\bibinfo{author}{\bibfnamefont{S.~W.} \bibnamefont{{Barwick}}}
  \bibnamefont{et~al.} (\bibinfo{collaboration}{HEAT}),
  \bibinfo{journal}{\apjl} \textbf{\bibinfo{volume}{482}},
  \bibinfo{pages}{L191+} (\bibinfo{year}{1997}), \eprint{astro-ph/9703192}.

\bibitem[{\citenamefont{Aguilar et~al.}(2007)}]{Aguilar:2007yf}
\bibinfo{author}{\bibfnamefont{M.}~\bibnamefont{Aguilar}} \bibnamefont{et~al.}
  (\bibinfo{collaboration}{AMS-01}), \bibinfo{journal}{Phys. Lett.}
  \textbf{\bibinfo{volume}{B646}}, \bibinfo{pages}{145} (\bibinfo{year}{2007}),
  \eprint{astro-ph/0703154}.

\bibitem[{\citenamefont{Alcaraz et~al.}(2000)}]{Alcaraz:2000bf}
\bibinfo{author}{\bibfnamefont{J.}~\bibnamefont{Alcaraz}} \bibnamefont{et~al.}
  (\bibinfo{collaboration}{AMS}), \bibinfo{journal}{Phys. Lett.}
  \textbf{\bibinfo{volume}{B484}}, \bibinfo{pages}{10} (\bibinfo{year}{2000}).

\bibitem[{\citenamefont{{Adriani} et~al.}(2008)}]{PAMELA_frac}
\bibinfo{author}{\bibfnamefont{O.}~\bibnamefont{{Adriani}}}
  \bibnamefont{et~al.} (\bibinfo{collaboration}{PAMELA})
  (\bibinfo{year}{2008}), \eprint{0810.4995}.

\bibitem[{\citenamefont{{Casadei} and {Bindi}}(2004)}]{2004ApJ...612..262C}
\bibinfo{author}{\bibfnamefont{D.}~\bibnamefont{{Casadei}}} \bibnamefont{and}
  \bibinfo{author}{\bibfnamefont{V.}~\bibnamefont{{Bindi}}},
  \bibinfo{journal}{\apj} \textbf{\bibinfo{volume}{612}}, \bibinfo{pages}{262}
  (\bibinfo{year}{2004}).

\bibitem[{\citenamefont{{Bergstrom} et~al.}(2008)\citenamefont{{Bergstrom},
  {Bringmann}, and {Edsjo}}}]{Bergstrom:2008gr}
\bibinfo{author}{\bibfnamefont{L.}~\bibnamefont{{Bergstrom}}},
  \bibinfo{author}{\bibfnamefont{T.}~\bibnamefont{{Bringmann}}},
  \bibnamefont{and} \bibinfo{author}{\bibfnamefont{J.}~\bibnamefont{{Edsjo}}}
  (\bibinfo{year}{2008}), \eprint{0808.3725}.

\bibitem[{\citenamefont{Bertone et~al.}(2005)\citenamefont{Bertone, Zentner,
  and Silk}}]{Bertone:2005xz}
\bibinfo{author}{\bibfnamefont{G.}~\bibnamefont{Bertone}},
  \bibinfo{author}{\bibfnamefont{A.~R.} \bibnamefont{Zentner}},
  \bibnamefont{and} \bibinfo{author}{\bibfnamefont{J.}~\bibnamefont{Silk}},
  \bibinfo{journal}{Phys. Rev.} \textbf{\bibinfo{volume}{D72}},
  \bibinfo{pages}{103517} (\bibinfo{year}{2005}), \eprint{astro-ph/0509565}.

\bibitem[{\citenamefont{Brun et~al.}(2007)\citenamefont{Brun, Bertone, Lavalle,
  Salati, and Taillet}}]{Brun:2007tn}
\bibinfo{author}{\bibfnamefont{P.}~\bibnamefont{Brun}},
  \bibinfo{author}{\bibfnamefont{G.}~\bibnamefont{Bertone}},
  \bibinfo{author}{\bibfnamefont{J.}~\bibnamefont{Lavalle}},
  \bibinfo{author}{\bibfnamefont{P.}~\bibnamefont{Salati}}, \bibnamefont{and}
  \bibinfo{author}{\bibfnamefont{R.}~\bibnamefont{Taillet}},
  \bibinfo{journal}{Phys. Rev.} \textbf{\bibinfo{volume}{D76}},
  \bibinfo{pages}{083506} (\bibinfo{year}{2007}), \eprint{0704.2543}.

\end{thebibliography}
\end{document}